\documentclass[twocolumn,amsmath,amssymb,nofootinbib,preprintnumbers,showpacs]{revtex4}

\usepackage[pdftex]{graphicx}
\usepackage{epstopdf}

\usepackage{amsmath}
\usepackage{amssymb}
\usepackage{subfigure}
\usepackage{hyperref}
\usepackage{url}
\usepackage{xcolor}
\usepackage{color}
\definecolor{amaranth}{rgb}{0.9, 0.17, 0.31}
\definecolor{purple(munsell)}{rgb}{0.62, 0.0, 0.77}
\definecolor{americanrose}{rgb}{1.0, 0.01, 0.24}
\definecolor{palatinateblue}{rgb}{0.15, 0.23, 0.89}
\definecolor{royalblue(web)}{rgb}{0.25, 0.41, 0.88}
\definecolor{hanpurple}{rgb}{0.32, 0.09, 0.98}
\definecolor{beaublue}{rgb}{0.74, 0.83, 0.9}
\definecolor{carminered}{rgb}{1.0, 0.0, 0.22}
\definecolor{brightpink}{rgb}{1.0, 0.0, 0.5}
\hypersetup{ linktoc=all,
    colorlinks, linkcolor={palatinateblue},
    citecolor={brightpink}, urlcolor={palatinateblue}
}

\def\sideremark#1{\ifvmode\leavevmode\fi\vadjust{\vbox to0pt{\vss
 \hbox to 0pt{\hskip\hsize\hskip1em
 \vbox{\hsize2cm\tiny\raggedright\pretolerance10000
 \noindent #1\hfill}\hss}\vbox to8pt{\vfil}\vss}}}%
\begin{document}
   \title{Accelerating Plasma Mirrors to Investigate Black Hole Information Loss Paradox}
   \author{Pisin Chen$^{1,2}$\footnote{pisinchen@phys.ntu.edu.tw},
   Gerard Mourou$^{3}$\footnote{gerard.mourou@polytechnique.edu},
      }

   \affiliation{%
   ~\\
$^{1}$Leung Center for Cosmology and Particle Astrophysics \& Department of Physics and Graduate Institute of Astrophysics, National Taiwan University, Taipei 10617, Taiwan\\
$^{2}$Kavli Institute for Particle Astrophysics and Cosmology, SLAC National Accelerator Laboratory, Stanford University, CA 94305, U.S.A.\\
$^{3}$IZEST, Ecole Polytechnique, 91128 Palaiseau Cedex, France
}%


\begin{abstract}
The question of whether Hawking evaporation violates unitarity, and therefore results in the loss of information, remains unresolved since Hawking's seminal discovery. So far the investigations remain mostly theoretical since it is almost impossible to settle this paradox through direct astrophysical black hole observations. Here we point out that relativistic plasma mirrors can be accelerated drastically and stopped abruptly by impinging intense x-ray pulses on solid plasma targets with a density gradient. This is analogous to the late time evolution of black hole Hawking evaporation. A conception of such an experiment is proposed and a self-consistent set of physical parameters is presented. Critical issues such as how the black hole unitarity may be preserved can be addressed through the entanglement between the analog Hawking radiation photons and their partner modes. 
\end{abstract}

\date{Feburary 28, 2016}

\maketitle

The question of whether Hawking evaporation \cite{Hawking 1974} violates unitarity, and therefore results in the loss of information \cite{Hawking 1978}, remains unresolved since Hawking's seminal discovery. Proposed solutions range from black hole complementarity \cite{Susskind} to firewalls \cite{AMPS, AMPSS} (see, for example, \cite{Mathur, Chen 2014} for a recent review and \cite{Chen 2015} for a counter argument). So far the investigations remain mostly theoretical since it is almost impossible to settle this paradox through direct astrophysical observations, as typical stellar size black holes are cold and young. Here we point out that the Hawking evaporation at its late stage can be mimicked by accelerating plasma mirrors based on state-of-the-art laser and nano-fabrication technologies. We show that a relativistic plasma mirror \cite{Bulanov, Maumova, Bulanov 2013} induced by an intense x-ray beam traversing a solid plasma target with an increasing density and a sharp termination can be accelerated drastically and stopped abruptly. Critical issues such as how the black hole unitarity would be preserved can in principle be addressed.

There have been proposals for laboratory investigations of the Hawking effect, including sound waves in moving fluids \cite{Unruh 1980},  electromagnetic waveguides \cite{waveguide1}, traveling index of refraction in media \cite{Yablonovitch}, Bose-Einstein condensates \cite{Steinhauer 2014}, and electrons accelerated by intense lasers \cite{Chen 1999}. In particular, Ref.[15] reported on the observation of quantum Hawking radiation and its entanglement. 
It has long been recognized that accelerating mirrors can mimic black holes and emit Hawking-like thermal radiation \cite{Fulling-Davies1}. That accelerating mirrors can also address the information loss paradox was first suggested by Wilczek \cite{Wilczek}. 

As is well-known, the notion of black hole information loss is closely associated with quantum entanglement. In order to preserve the ``black hole unitarity", Wilczek argued, based on the moving mirror model, that the partner modes of the Hawking particles would be trapped by the horizon until the end of the evaporation, where they would be released and the black hole initial pure state recovered with essentially zero cost of energy. More recently, Hotta et al. \cite{Hotta} confirmed that the released partner modes are simply indistinguishable from the zero-point vacuum fluctuations. On the other hand, there is also the notion that these partner modes would be released in a burst of energy, for example, in the Bardeen model \cite{Bardeen}. It is even more important to determine how the black hole information is retrieved. Does it follow the Page curve \cite{Page 1993}, or some alternative scenario \cite{Hotta-Sugita}? This would critically impact on various conjectures such as firewalls that assume certain scenario for the entanglement entropy.
The observation of either a burst of radiation or zero-point fluctuations, and the measurement of the entanglement between these modes and the Hawking particles and its evolution, should help to shed much lights on the black hole information loss paradox.



Plasma wakefield acceleration for high energy particles driven by lasers \cite{Tajima} or particle beams \cite{Chen 1985} has been a subject of worldwide pursuit. In the nonlinear regime of plasma perturbations, the plasma wakefield will undergo ``wave-breaking" that results in a huge pileup of density perturbation, like a tsunami. It has been proposed that this feature can provide yet another salient utility, a highly relativistic plasma mirror, that can reflect and Lorentz-boost a witness optical laser pulse to turn it into an even more compressed x-ray beam \cite{Bulanov, Maumova, Bulanov 2013}. 

By the conservation of energy, the driving laser pulse must lose its energy by exciting the wakefield. Since the photon number is nearly conserved in such system, the energy loss is manifested by the frequency redshift, which causes the slow-down of the laser \cite{Sprangle, Esarey 1990, Spitkovsky}. 
To counter this {\it natural} tendency of deceleration, one can envision an {\it artificial} inducement of acceleration of the laser by traversing a plasma with increasing density. Such accelerating mirror plays the role of the black hole center, while the mirror's asymptotic null ray serves as the equivalent horizon. Additionally, terminating the plasma target sharply would result in a sudden stoppage of the mirror motion, which would mimic the end-life of black hole \cite{Wilczek}, independent of whether it would evaporate entirely or end with a Planck-size remnant \cite{Adler 2001,Chen 2014}.

In the adiabatic limit, the laser-plasma interaction can be described in the comoving coordinates $\chi\equiv x-v_gt$ and $\tau=t$ by the following nonlinear coupled equations:
\begin{eqnarray}\label{LaserPlasmaInt}
\Big[\frac{2}{c}\frac{\partial}{\partial\chi}-\frac{1}{c^2}\frac{\partial}{\partial\tau}\Big]\frac{\partial a}{\partial \tau}&=&k_{p0}^2\frac{a}{1+\phi}, \\
\frac{\partial^2\phi}{\partial\chi^2}&=&-\frac{k_{p0}^2}{2}\Big[ 1-\frac{(1+a)^2}{(1+\phi)^2}\Big],
\end{eqnarray}
where $\phi$ and $a$ are the (normalized dimensionless) scalar and vector potentials of the laser, and $k_{p0}=\omega_{p0}/c=\sqrt{4\pi r_e n_{p0}}$. Here $r_e=e^2/mc^2$ is the classical electron radius, $n_{p0}$ is the ambient uniform plasma density. According to the {\it principle of wakefield}, the phase velocity of the plasma wakefield equals the group velocity of the driving laser or particle beam. For uniform plasmas, $v_{ph0}=v_{g0}=\eta_0 c$, where the refractive index $\eta_0=\sqrt{1-(\omega_{p0}^2/\omega_0^2)/(1+\phi)}\leq 1$. In the nonlinear regime, where the laser strength parameter $a_{_L0} > 1$, the density perturbation approaches a delta function, but periodic nonetheless. As discussed above, the laser loses its energy through frequency redshift \cite{Esarey 1990, Spitkovsky}:
\begin{eqnarray}\label{redshift}
\frac{\partial\omega}{\partial \chi}=-\frac{\omega_{p0}^2}{2\omega_0}\frac{\partial}{\partial \chi}\Big(\frac{1}{1+\phi}\Big).
\end{eqnarray}
As the frequency decreases, $\eta_0$ decreases as well, and so the laser slows down as it traverses the plasma.

Now we model the laser intensity variation along the pulse as a sine function \cite{Sprangle}:
\begin{eqnarray}\label{lasershape}
a_{L}(\chi)=
\begin{cases}
a_{L 0}\sin(\pi\chi/L), \quad\quad -L<\chi\leq 0, \\
0, \quad\quad\quad\quad\quad\quad\quad\quad {\rm otherwise},
\end{cases}
\end{eqnarray}
Under the assumption that $\phi\ll 1$, which is satisfied even if $a_{_L0}>1$ as long as $L\ll \lambda_p$ \cite{Sprangle}, the solution to Eq.(2) for the ``slowly varying part" of the scalar potential is
\begin{eqnarray}\label{scalarpotential}
\phi\simeq \frac{a_{L 0}^2 k_{p0}^2}{8} \Big\{\chi^2-2\Big(\frac{L}{2\pi}\Big)^2 \Big[1-\cos(2\pi\chi/L)\Big] \Big\}.
\end{eqnarray}
The variation of $\phi$ along the laser pulse is then
\begin{eqnarray}\label{scalarpotential}
\frac{\partial\phi}{\partial\chi}\simeq \frac{a_{L0}^2k_{p0}^2}{4}\big[\chi-\frac{L}{\pi}\sin\big(\frac{2\pi\chi}{L}\big)\big], \quad -L\leq \chi \leq 0.
\end{eqnarray}
We see that $\partial\phi/\partial\chi$ is negative definite. Therefore, $\partial\omega/\partial \chi$, as defined in Eq.(3), is also negative definite. 

The plasma dispersion relation and phase velocity under nonuniform density have been investigated and confirmed through computer simulations by Lobet et al. \cite{Lobet 2013}. By definition, the eikonal of the plasma wave, $\theta(x,t)$, satisfies the relationships $\omega_p=-\partial\theta/\partial t$ and $k_p=\partial \theta/\partial x,$ and has the cross differentiation property $\partial^2\theta/\partial t\partial x = \partial^2\theta/\partial x\partial t$. As a consequence, we have 
\begin{eqnarray}\label{refractiveindex}
\frac{\partial k_p}{\partial t}=-\frac{\partial \omega_p}{\partial x}.
\end{eqnarray}
As plasma oscillations are a collective effect, the minimum length scale for the response to density variation is the plasma wavelength, $\lambda_p$. Let the scale length of the density variation be $D$. If $\lambda_p\ll D$, then the wave number $k_p(x+\Delta x)$ can be related to that at $x$ through the Taylor expansion: $k_{p}(x+\Delta x)=k_{p}(x)+(\partial k_p/\partial t)_x\Delta t=k_{p}(x)-(\partial \omega_p/\partial x)_x\Delta t$, where $\Delta t \ll D/c$. Substituting it in the phase velocity, we arrive at 
\begin{eqnarray}\label{refractiveindex}
v_{ph}(x+\Delta x)\simeq v_{ph}(x)\Big(1+\frac{\partial \omega_p}{\partial x}\frac{\Delta t}{k_{p}(x)}\Big). 
\end{eqnarray}
Identifying $v_{ph}(x+\Delta x)-v_{ph}(x)=\Delta v_{ph}(x)$ and dividing both sides by $\Delta t$, we find $d v_{ph}/dt=v_{ph}[\partial \omega_p/\partial x)/k_p]$. As $\Delta x=v_{ph}\Delta t$ and $\omega_p=v_{ph}k_p$ locally, we substitute $\Delta t/k_p$ with $\Delta x/\omega_p$ and find
\begin{eqnarray}\label{refractiveindex}
v_{_M}\equiv v_{ph}(x)\simeq v_{ph0}\exp\Big\{\frac{\partial \omega_p}{\partial x}\frac{x}{\omega_p}\Big\}.
\end{eqnarray} 
We see that if $\partial\omega_p/\partial x>0$, then the velocity of the plasma mirror would increase in time.

Since $d/dt=\partial/\partial t+v_g\partial/\partial x$, the acceleration of the plasma mirror is 
\begin{eqnarray}\label{scalarpotential}
\ddot{x}_{_M}&=&\frac{c}{2\eta_0}\Big[v_g\Big(1+\frac{\omega_{p0}^2}{\omega^2}\Big)\frac{\omega_{p0}^2}{\omega^2}\frac{\partial}{\partial x}\frac{1}{1+\phi}\Big]\exp\Big\{\frac{\partial \omega_p}{\partial x}\frac{x}{\omega_{p}}\Big\}  \cr
&+&c\eta_0 v_g\Big(\frac{\partial \omega_p}{\partial x}\frac{1}{\omega_{p}}+\frac{\partial^2 \omega_p}{\partial x^2}\frac{x}{\omega_{p}}\Big)\exp\Big\{\frac{\partial \omega_p}{\partial x}\frac{x}{\omega_{p}}\Big\} .
\end{eqnarray}
The first term corresponds to the natural deceleration while the second term the artificial acceleration. As a prerequisite for laser propagation in a plasma, $\omega_p^2/\omega_0^2 \ll 1$, thus the gradient induced acceleration can in general dominate over the natural deceleration. From here on we will ignore the first term, which also implies that $v_g=v_{g0}=v_{ph0}$.

For a moving mirror to mimic the physics of black hole evaporation, it is necessary that its trajectory approaches the null ray asymptotically, which acts as a horizon. Guided by that, we invoke the following target density,
\begin{eqnarray}\label{plasmadensity}
n_p(x)=
\begin{cases}
n_{p0}(1+x/D)^{2(1-\eta_0)}, \quad\quad 0\leq x\leq X, \\

0, \quad\quad\quad\quad\quad\quad\quad\quad\quad\quad {\rm otherwise},
\end{cases}
\end{eqnarray}
Accordingly, $\omega_p(x)=\omega_0(1+x/D)^{(1-\eta_0)}$ and, based on Eq.(9), the velocity indeed approaches the speed of light:
\begin{eqnarray}
\frac{v_{ph}}{c}=1-\frac{1}{2}\frac{\omega_{p0}^2}{\omega_0^2}\Big[1-\frac{x/D}{1+x/D}\Big]+\mathcal{O}\Big(\frac{\omega_{p0}^4}{\omega_0^4}\Big).
\end{eqnarray}
Keeping all orders, the mirror acceleration is
\begin{eqnarray}\label{refractiveindex}
\ddot{x}_{_M}(x) \simeq \frac{(1-\eta_0)c^2}{D(1+x/D)^2}\exp\Big\{\frac{(1-\eta_0)x/D}{1+x/D}\Big\},
\end{eqnarray}
and the corresponding analog Hawking temperature \cite{Davies 1974} is then
\begin{eqnarray}\label{refractiveindex}
k_{_B}T_{_H}(x)\simeq\frac{\hbar c}{4\pi D}\frac{\omega_{p0}^2}{\omega_0^2}\frac{1}{(1+x/D)^2}\exp\Big\{\frac{(1-\eta_0)x/D}{1+x/D}\Big\}.
\end{eqnarray}
In order that the Hawking radiation be truly thermal, the change of the Hawking temperature must be sufficiently adiabatic such that the vacuum fluctuating photons can ``respond" to the instantaneous temperature. The characteristic time of Hawking temperature variation is $v_{ph}/D\sim c/D$ while that for the fluctuating photons is the inverse of the characteristic frequency of the Hawking spectrum before the redshift, $4\gamma^2\omega_{H}=4\gamma^2 k_BT_H/\hbar$, where $\gamma^2=1-v_{ph}^2/c^2\simeq \omega_0^2/\omega_p^2$. So the adiabatic condition can be cast as $A\equiv (4\gamma^2\omega_H)^{-1}(c/D)<1$. We find, for our specific density profile,  $A\sim 2\pi/(1+x/D)^2\leq 1$ for $x\geq 1.5D$. We conclude that the adiabatic condition is satisfied after the mirror traverses the first couple characteristic distances inside the target. 

Figure 1 shows the worldline of the accelerating plasma mirror and the spacetime evolution of the entangled vacuum fluctuating pairs. The partner modes of the Hawking particles are temporarily trapped by the horizon and would presumably be released when the mirror stops abruptly. Depending on different theories, there would be either a burst of real particles, or zero-point vacuum fluctuations \cite{Wilczek, Hotta}. The energies of the photons of the burst, if exists, should be a factor $4\gamma^2$ higher than that of the Hawking particles, which suffered a redshift as they bounced off from the receding plasma mirror. Through the measurement of the two-point correlation function between the Hawking particles and either the burst of energy or zero-point fluctuations, the entanglement entropy of the system as a function of time can be deduced. 

\begin{figure}
\includegraphics[width=\columnwidth]{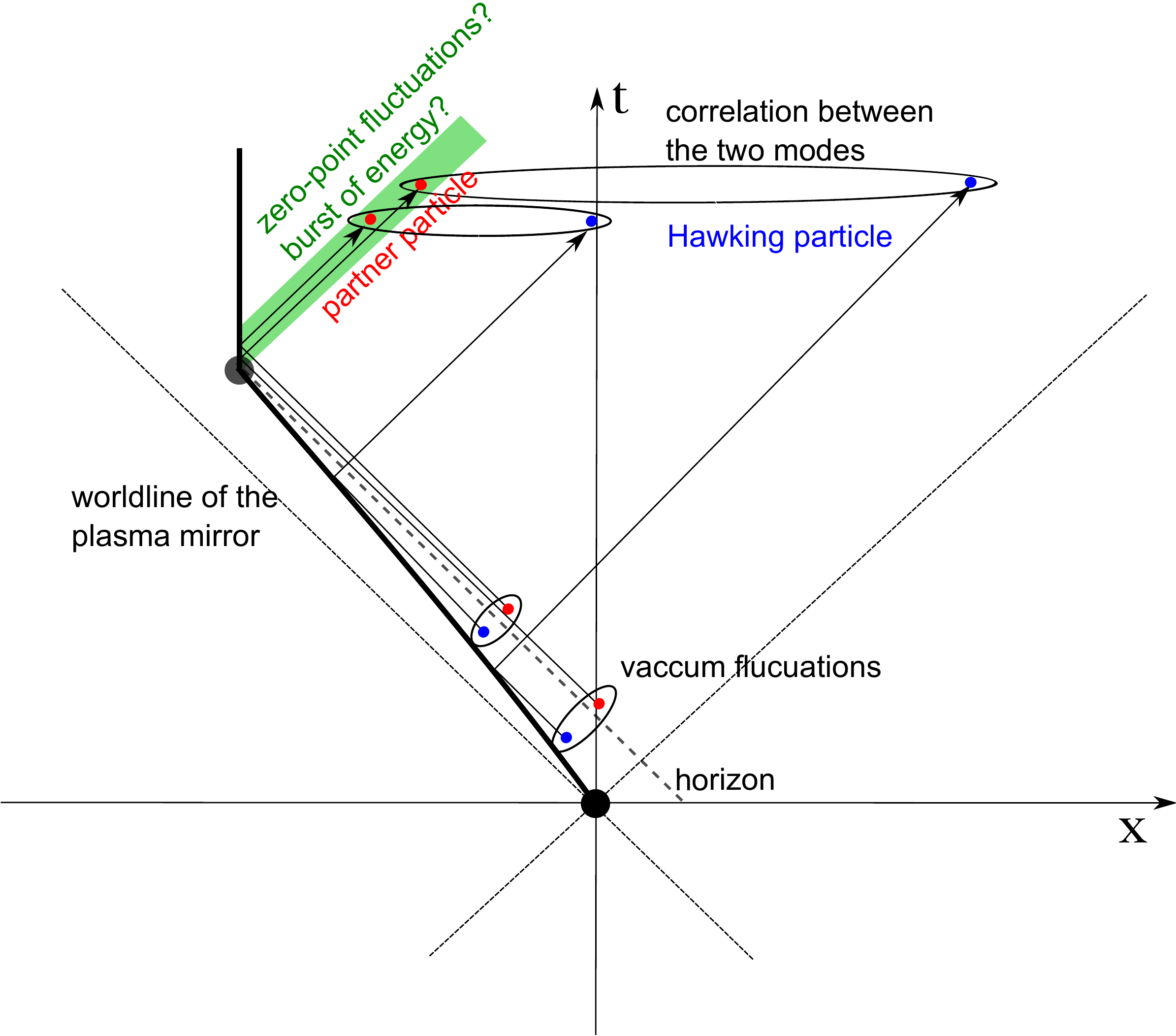}
\caption{The worldline of an accelerating relativistic plasma mirror and its relation with vacuum fluctuations around the horizon. In particular, the entanglement between the Hawking particles (blue) emitted at early times and their partner particles (red) collected at late times is illustrated. The green strip represents either a burst of energy or zero-point fluctuations emitted when the acceleration stops abruptly.}
\label{f2}
\end{figure}

\begin{figure*}[htp]
\includegraphics[clip, width=6 in]{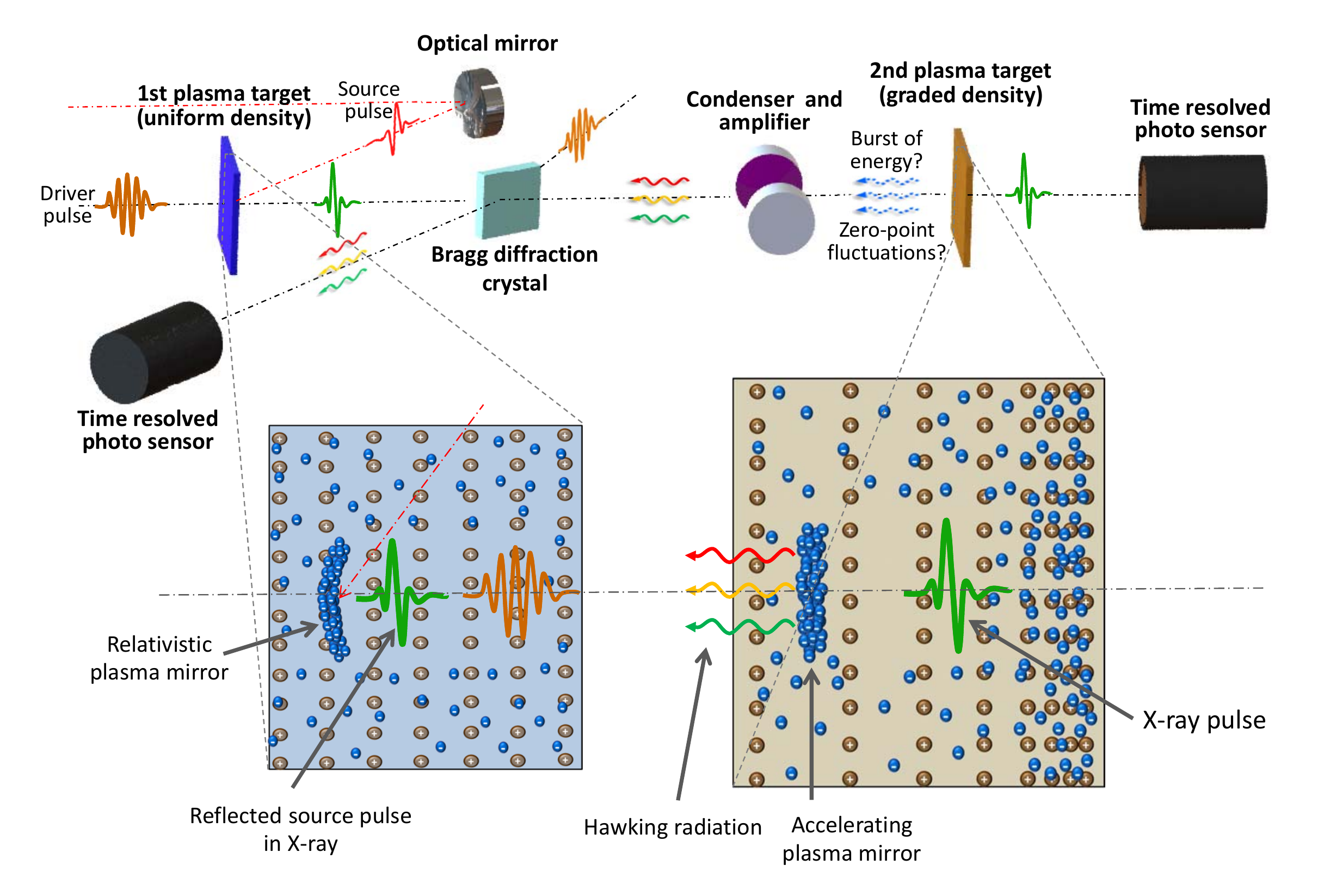}
\caption{A schematic diagram of the proposed analog black hole experiment. The first, gaseous and uniform plasma target is used to prepare a high intensity x-ray pulse. The x-ray pulse will induce an accelerating plasma mirror due to the increasing plasma density in the second target. As the mirror stops abruptly, it will release either a burst of energy or zero-point fluctuations. The correlation function between either of these signals and the Hawking photons is measured upstream.}
\end{figure*}

Figure 2 is a schematic diagram for our proposed experiment. In our conception, a driver pulse from an optical laser traverses the 1st (gaseous and uniform) plasma target, which creates a relativistic plasma mirror with a concave transverse density distribution. A source pulse, prepared by the same laser, is reflected by the plasma mirror with frequency increased by a factor $4\gamma^2$, where $\gamma$ is the Lorentz factor of the 1st mirror. This x-ray pulse will pass through a Bragg diffraction crystal to arrive at the 2nd plasma target, which is solid with graded density, presumably fabricated via nano-technology. The driver pulse, on the other hand, can be diffracted to a different path. 

This second plasma mirror accelerates due to the density gradient and emits the high-frequency part of the analog Hawking radiation, which propagates in the backward direction. 
At the time when the mirror arrives at the back-end, it stops abruptly. At this point all the partner modes would be suddenly released and travel backwards. Upstream of the 2nd target, the supposed zero-point fluctuations will be measured by condensers and amplifiers, while the Hawking and the partner particles (if real), which are sufficiently lower in frequency than that of the x-ray, will be Bragg diffracted to a time-resolved photo-sensor. The time resolution should be much finer than the penetration time, say femtosecond, such that the final burst of partner particles can be distinguished from the Hawking photons.

As a numerical example, let us consider a 10 PetaWatt single-cycle green laser to produce the x-ray pulse. A minor fraction of this power is used as the driver pulse that impinges the 1st target with a gas density of $\sim 2\times 10^{19}{\rm cm}^{-3}$, which, by design, induces a concave, constant velocity relativistic plasma mirror with the Lorentz factor $\gamma=\omega_d/\omega_p \sim 10$. The source pulse carrying essentially 10 PW, with $5 \times 10^{19}$ photons per pulse, counter-propagates to collide with the mirror. After reflection, it is Lorentz boosted to $\hbar \omega_x = 4\gamma^2\omega_s\sim 1 {\rm keV}$. This corresponds to a wavelength $\lambda_x\sim 1.2 {\rm nm}$. The reflectivity of a plasma mirror depends on the laser frequency and the plasma density \cite{Bulanov2013}. For the parameters of the 1st mirror, it is $Y\sim10^{-5}$ \cite{Kando}. Then the number of photons in the x-ray pulse is $N_x=5\times 10^{14}$. Let the depth of focus of the x-ray be much larger than the 2nd target thickness $X$ so as to maintain a near-constant pulse radius $R\sim 330\lambda_x\sim 400{\rm nm}$. Then the x-ray strength parameter is $a_x=eE/mc\omega_x\sim 2 >1$, which is sufficient to excite nonlinear plasma wakefields for the accelerating mirror.

There are four key length parameters in our proposed experiment, which, due to various constraints, follow the inequality: $\lambda_x \ll \lambda_p(x)\ll D \ll X$. Let the initial plasma density be $n_{p0}=1.3\times 10^{25} {\rm cm}^{-3}$. This corresponds to $\lambda_{p0}=9.3 {\rm nm}$ and $\omega_{p0}^2/\omega_0^2\sim 0.64$. Let $D=100 {\rm nm}$ and $X=5D$. 
Then at the back-end of the target, the density reaches $n_{pf}\sim 4.1\times 10^{25} {\rm cm}^{-3}$. The Hawking temperature decreases from $\sim 0.1{\rm eV}$ at $x=0$ to $0.004 {\rm eV}$ at $x=5D$. Limited by the size of the mirror, $2R_x\sim 800 {\rm nm}$, the Hawking spectrum below 1.8 eV would be cutoff. Since, dictated by the adiabatic condition, only the high frequency part of the Hawking spectrum can be emitted by the mirror, these two constraints roughly coincide accidentally. It can be verified \cite{Bulanov2013} that the reflectivity for the 2nd mirror approaches unity, thanks to the high density of the solid target. In the $Y\sim 1$ limit, the reflectivity becomes independent of the reflecting photon frequency, which helps to preserve the exponential tail of the Hawking spectrum.


For this experiment to render meaningful result, the Hawking signals must compete successfully against various inevitable background photons at comparable energies and signatures. The conventional Compton scattering between the intense x-ray photons and the target electrons is not a concern as   such photons are essentially forward moving while our signals move backward. 

The more dangerous backgrounds would be entangled double-photon emissions via either violently accelerated (by the intense x-ray pulse) electrons \cite{Schutzhold 2008}, or double-photon Compton back-scatterings as a result of multi-photon laser-electron interactions \cite{Lotstedt 2009}. In the former case the pair photons move forward, governed by the x-ray ponderomotive force, and is opposite to that of our signals. As for the latter case, a main difference is the temporal separations between the two photons in the pair. For multi-photon Compton scatterings, the separation time is the duration of the single-cycle x-ray laser pulse, which in our case is $\lambda_x\sim$ 1nm. In contrast, the partner modes in our experiment are trapped by the accelerating mirror's horizon until it stops, and so their time separation from the Hawking modes is of the order of the target thickness, or $D\sim {\rm 100nm}\gg \lambda_x$. 

We emphasize that this ``nonlocal" separation between the two entangled photons is a salient feature of our proposed experiment. Another character of our experiment is the distinct energy separation between the pair photons by a factor $4\gamma^2$, which should help to further differentiate them from those double photon events described above.

Moving mirrors can provide additional utilities for investigating black hole physics. As was pointed out by Wilczek \cite{Wilczek}, a rapidly receding mirror has a dynamical effect that mimics the redshift due to the spacetime distortion near the surface of the black hole. In addition, having a finite mass, the plasma mirror should recoil upon thermal emissions, which would presumably provide a model for the intrinsic black hole entropy \cite{Wilczek}.

We are grateful to S. Bulanov, M. Hotta, J. Nam, Y. C. Ong, D. Page, W. Unruh, and D-h. Yeom for many helpful comments and suggestions. PC is supported by Ministry of Science and Technology (MOST) of Taiwan, and the Leung Center for Cosmology and Particle Astrophysics (LeCosPA), National Taiwan University.

\end{document}